\documentclass[conference,letterpaper]{IEEEtran}

\usepackage[utf8]{inputenc} 
\usepackage[T1]{fontenc}
\usepackage{url}
\usepackage{ifthen}
\usepackage{cite}
\usepackage[cmex10]{amsmath} 

\usepackage{algorithm}
\usepackage{algpseudocode}
\usepackage{amsmath}
\usepackage{hyperref} 
\usepackage{amssymb}
\usepackage{graphicx,color,epsfig,rotating}

\usepackage{color}

\newcommand{\red}{\color{red}}

\long\def\comment#1{}

\newfont{\bbb}{msbm10 scaled 700}

\newfont{\bb}{msbm10 scaled 1100}

\newcommand{\Uc}{{\cal U}}

\newcommand{\csf}{{\sf c}}

\newcommand{\qsf}{{\sf q}}

\newcommand{\Ksf}{{\sf K}}
\newcommand{\Lsf}{{\sf L}}

\newcommand{\Nsf}{{\sf N}}

\newcommand{\Rsf}{{\sf R}}

\newcommand{\Tsf}{{\sf T}}
\newcommand{\Usf}{{\sf U}}

\newcommand{\Ysf}{{\sf Y}}

\newcommand{\be}{\begin{equation}}
\newcommand{\ee}{\end{equation}}
\newcommand{\bea}{\begin{eqnarray}}
\newcommand{\eea}{\end{eqnarray}}
\newtheorem{thm}{Theorem}

\begin{document}
\title{Multi-message Secure Aggregation with Demand Privacy} 



\author{
\IEEEauthorblockN{%
Chenyi Sun\IEEEauthorrefmark{1},
Ziting Zhang\IEEEauthorrefmark{1},
Kai Wan\IEEEauthorrefmark{1},
Giuseppe Caire\IEEEauthorrefmark{2}
}
\IEEEauthorblockA{\IEEEauthorrefmark{1}Huazhong University of Science and Technology, 430074  Wuhan, China, \\ \{chenyi\_sun, ziting\_zhang, kai\_wan\}@hust.edu.cn}%
 \IEEEauthorblockA{Technische Universit\"at Berlin, 10587 Berlin, Germany,   caire@tu-berlin.de}%
}

\maketitle


\begin{abstract}
   This paper considers a  multi-message secure aggregation with demand privacy problem,  in which a server aims to compute $\sf K_c\geq 1$ linear combinations of local inputs from $\Ksf$ distributed users. The problem addresses two tasks: (1) security, ensuring that the server can only obtain the desired linear combinations without any else information about the users' inputs, and (2) privacy, preventing users from learning about the server's computation task.  In addition, the effect of user dropouts is considered, where at most $\Ksf-\Usf$ users can drop out and the identity of these users cannot be predicted in advance. We propose two schemes for $\sf K_c=1$ and $2\leq {\sf K_c}< \Usf$, respectively.
  For $\sf K_c=1$, we introduce multiplicative encryption of the server's demand using a random variable, where users share coded keys offline and transmit masked models in the first round, followed by aggregated coded keys in the second round for task recovery. For $2\leq {\sf K_c}< \Usf$, we use robust symmetric private computation to recover linear combinations of keys in the second round. The objective is to minimize the number of symbols sent by each user during the two rounds. Our proposed schemes have achieved the optimal rate region when $\sf K_c=1$ and the order optimal rate (within $2$) when $2\leq {\sf K_c}< \Usf$.
\end{abstract}


\section{Introduction}
\label{section: intro}
Federated learning allows distributed users to collaboratively train a global model under the coordination of a central server \cite{mcmahan2016federated}. Instead of sharing raw data, users perform local model training on their own datasets and send only model updates (e.g. gradients or weights) to the server, which aggregates these updates to a global model iteratively \cite{bonawitz2019towards}.

However, the updates in federated learning can still reveal some information about users' local data \cite{kairouz2021advances}. To further enhance security, secure aggregation was introduced in \cite{bonawitz2017practical}, which guarantees that the server can only get the aggregated updates without gaining any additional information about users' data.
The first information theoretic formulation of secure aggregation was proposed in \cite{zhao2022information}. To address the effect of user dropouts, two rounds of transmission are necessary. In the first round, each user encrypts its local model with stored keys, and in the following second round, surviving users further send messages about keys according to the set of non-dropped users in the first round. The objective is to characterize the region of all achievable communication rate tuples $(\Rsf_1,\Rsf_2)$, where $\Rsf_i$ represents the largest number of symbols per input symbol transmitted in the $i$-th round message. The scheme in \cite{zhao2022information} has achieved the capacity region $\{(\Rsf_1,\Rsf_2):\Rsf_1\geq 1, \Rsf_2\geq 1/\Usf\}$,  where $\Usf$ represents the minimum number of non-dropped users.
The secure aggregation scheme proposed in \cite{so2022lightsecagg} achieves the same capacity as in \cite{zhao2022information}, but requires significantly fewer keys by decoupling the key generation process from the set of non-dropped users in the first round.

In this paper, we focus on a new formulation of secure aggregation, {\it multi-message secure aggregation with demand privacy}. Compared to the information theoretic secure aggregation problem in~\cite{zhao2022information}, the considered problem has two main differences: (i) the server requests $\Ksf_\csf \geq 1$ linear combinations of the 
users' local models, while in~\cite{zhao2022information} $\Ksf_\csf$ is set to $1$\footnote{\label{foot:coordinated learning}{Compared to single-task, multiple-task learning can leverage multiple results to enhance overall performance of model training \cite{zhang2021survey}. For instance, coordinated learning enables multiple tasks to collaborate by sharing some information and resources, thereby improving overall learning efficiency \cite{guestrin2002coordinated}. In this approach, tasks benefit from common representations, gradients, or shared components, allowing them to learn complementary features.}}; (ii) the constraint of demand privacy is considered, where the users cannot know any information about the coefficients in the linear combinations of models.

Demand privacy against distributed nodes was widely considered in the problem of Private information retrieval (PIR), introduced in \cite{chor1998private}. The problem aims to enable the server to retrieve one message out of all messages stored by distributed databases as efficiently as possible, without revealing the identity of the desired message. The information theoretic capacity of PIR 
was established in \cite{sun2017capacity}, 
which characterizes the maximum number of symbols of desired information that can be privately retrieved per downloaded symbol.
As a generalization of PIR, private computation (PC) extends the framework to include the linear computations of these messages \cite{sun2018capacity,mirmohseni2018private}. Interestingly, the capacity of the PC problem remains the same as that of the PIR problem. Considering the demand of multiple linear combinations of messages, a multi-message private computation scheme was proposed in~\cite{mmpc2024Gholami} by non-trivially combining the multi-message PIR scheme in~\cite{multimessage2018} and the PC scheme in~\cite{sun2018capacity}.
The original formulation of PC focuses solely on protecting the privacy of the server's computation task while neglecting the security of users' local data.
To address this limitation, symmetric PC is proposed \cite{sun2018spircapacity,zhu2022symmetric}. In the scheme proposed in \cite{zhu2022symmetric}, the server can retrieve a polynomial function of all messages stored in distributed databases without any database learning the identity of the requested computation. At the same time, the server obtains no information about the messages other than the desired result. 
The rate achieved by this scheme is strictly less than the capacity of the PC due to the additional constraints imposed by the security requirements. However, the formulated problem in this paper is not a direct extension of symmetric PC; this is because, we consider the two-round secure aggregation problem where each user holds an individual message.

\paragraph*{Main Contribution}
Besides the new problem formulation on multi-message secure aggregation with demand privacy, our main contribution is as follows. 
\begin{itemize}
    \item For the case of ${\sf \sf K_c}=1$, we propose a secure aggregation scheme with demand privacy by combining the multiplicative encryption on demand and the additive encryption on models. This scheme achieves the optimal communication rate tuple,
    which is equal to the capacity of the secure aggregation problem without privacy constraint in~\cite{zhao2022information}.
    \item For the case of $2\leq {\sf K_c}< \Usf$,  we propose a secure aggregation scheme by leveraging the symmetric PC scheme in \cite{zhu2022symmetric}. This scheme significantly improves upon the direct approach of repeating the first scheme ${\sf K_c}$ times. Specifically, its first-round rate is optimal while its second-round rate is order optimal within $2$.
\end{itemize}

\paragraph*{Notation Convention}
Calligraphic symbols represent sets. Vectors and matrices are denoted using bold symbols. The system parameters are represented in sans-serif font. $|\cdot|$ defines the cardinality of a set. The notation $[a : b]$ represents the range $\{a,a+1,\dots,b\}$ and $[n]$ denotes the set $[1 : n]$. $\mathbb{F}_{\qsf}$ represents a finite field with order ${\qsf}$. $\mathbf{A}_{m \times n}$ represents that the matrix $\mathbf{A}$ is of dimension $m \times n$ and rank($\mathbf{A}$) indicates the rank of matrix $\mathbf{A}$. $\mathbf{A}^\Tsf$ denote the transpose of $\mathbf{A}$. Entropies are calculated on the base ${\qsf}$, where ${\qsf}$ denotes the field size.

\section{System Model and Related Results}

\subsection{System Model}
The formulated $(\Ksf, \Usf, \Ksf_\csf)$ multi-message secure aggregation with demand privacy problem contains a server and $\Ksf$ non-colluding users, where $\Ksf\geq 2$. User $i\in[\Ksf]$ holds an input vector $W_i$ and the key $P_i$. 
The input vector $W_i$ contains $\Lsf$ i.i.d. uniform symbols over a finite field $\mathbb{F}_\qsf$, and is independent of each other. We assume that $\Lsf$ is large enough. 
Furthermore, the keys $(P_i)_{i\in[\Ksf]}$ are independent of the input vectors $(W_i)_{i\in[\Ksf]}$,
\begin{align*}  H\left(\left(W_i\right)_{i\in[\Ksf]},\left(P_i\right)_{i\in[\Ksf]}\right)=\sum_{i\in[\Ksf]}H\left(W_i\right)+H\left(\left(P_i\right)_{i\in[\Ksf]}\right).
\end{align*}


The server aims to compute a function of users' inputs, denoted as $g(W_1,\cdots,W_\Ksf)$. For simplicity, we focus on the case where $g(\cdot)$ is a linear mapping. Thus, the task function can be expressed as:
\begin{align}
    g(W_1,\cdots,W_\Ksf) &= \mathbf{F}
   \begin{bmatrix}
       W_1, \cdots, W_\Ksf
   \end{bmatrix}^\Tsf\notag\\
    &=
   \begin{bmatrix}
       a_{1,1} & \cdots & a_{1,\Ksf}\\
       \vdots & \ddots & \vdots\\
       a_{\Ksf_\csf,1} & \cdots & a_{\sf K_c,K}
   \end{bmatrix}
   \begin{bmatrix}
       W_1\\\vdots\\W_\Ksf
   \end{bmatrix},
\end{align}
where $\mathbf{F}$ is the coefficient matrix with dimension $\sf {K_c}\times K$, each element $a_{n,i}$ for $n\in[\Ksf_\csf], i\in[\Ksf]$ is uniformly i.i.d. over $\mathbb{F}_{\qsf}$.  Without loss of generality, we assume that $\mathbf{F}$ is a row-wise full-rank matrix; otherwise, we
can just reduce $\sf {K_c}$ and let the server demand linearly independent combinations. In addition, we assume that there does not exist any $W_i$ whose coefficients in the demand are all $0$; otherwise, the server does not need to call this user $i$ into the computation process. 


{\bf First round.} In order to compute the desired function, the server has sent the query $Q_{1,i}$ to each user $i\in[\Ksf]$. User $i$ then feedbacks a message $X_i$ to the server. 
Some users may drop during this process, and the set of surviving users after the first round is denoted as $\mathcal{U}_1$, where $\mathcal{U}_1 \subseteq [\Ksf]$ and $|\mathcal{U}_1|\geq \Usf$.\footnote{\label{foot:value of U}In this paper, we assume that $\Usf\leq \Ksf-1$; otherwise, no user drops and one-round transmission is enough. } 

{\bf Second round.} The server informs all surviving users of $\mathcal{U}_1$ and requests a second round of messages from them to compute $\left(\sum_{i\in\mathcal{U}_1}a_{n,i}W_i\right)_{n\in[\sf K_c]}$. In addition, according to $\mathcal{U}_1$, the server also sends another query $Q^{\Uc_1}_{2,i}$ to each user $i\in \mathcal{U}_1$.
User $i\in \mathcal{U}_1$ then sends $Y_{i}^{\mathcal{U}_1}$ to the server. 
A subset of users may still drop and the surviving users are denoted as $\mathcal{U}_2$, where $\mathcal{U}_2\subseteq \mathcal{U}_1 \subseteq [\Ksf]$ and $|\mathcal{U}_2|\geq \Usf$. Thus, the server receives $(Y_{i}^{\mathcal{U}_1})_{i\in \mathcal{U}_2}$.\par

{\it Decodability.}  After receiving all the messages in the two rounds of transmission, the server can compute the desired linear combinations of the users' inputs $\left(\sum_{i\in\mathcal{U}_1}a_{n,i}W_i\right)_{n\in[\sf K_c]}$ without error, i.e., 
\begin{align}
H\Biggl(&\left.\left(\sum_{i\in\mathcal{U}_1}a_{n,i}W_i\right)_{n\in[\sf K_c]}\right|(X_i)_{i\in\mathcal{U}_1}, (Q_{1,i})_{i\in[\Ksf]}, \nonumber\\
&(Q^{\mathcal{U}_1}_{2,i})_{i\in\mathcal{U}_1}, (Y_{i}^{\mathcal{U}_1})_{i\in\mathcal{U}_2}\Biggr)= 0. \label{eq:decodability}
\end{align}

{\it Security.} The server cannot obtain any additional information about $(W_i)_{i\in[\Ksf]}$ beyond that contained
in $\left(\sum_{i\in{\mathcal{U}_1}}a_{n,i}W_i\right)_{n\in[\sf K_c]}$, i.e.,  
\begin{align}
    &I\Biggl(\left.\left(W_i\right)_{i\in[\Ksf]};\left(X_i\right)_{i\in{[\Ksf]}}, (Y_{i}^{\mathcal{U}_1})_{i\in\mathcal{U}_1}, (Q_{1,i})_{i\in[\Ksf]},(Q^{\mathcal{U}_1}_{2,i})_{i\in\mathcal{U}_1}\right| \nonumber\\
    &\left(\sum_{i\in \mathcal{U}_1}a_{n,i}W_i\right)_{n\in[\sf K_c]}\Biggr)=0.\label{eq:security}
\end{align}

{\it Privacy.} User $i\in[\Ksf]$ is unable to infer any information about the demand matrix $\mathbf{F}$, i.e.,   
\begin{align}  
   I\left(\mathbf{F};W_i,P_i, Q_{1,i}, Q^{\Uc_1}_{2,i}, X_i , Y_i^{\mathcal{U}_1}\right)=0,\forall i\in[\Ksf],\label{eq:privacy}  
\end{align}
where if $i\notin \Uc_1 $, $Y_i^{\mathcal{U}_1}=Q^{\Uc_1}_{2,i}=\emptyset$.

The two-round transmission rates are defined as follows,
\begin{align}
    \Rsf_1=\max_{i\in [\Ksf]} \frac{|X_i|}{\Lsf}, \ \ \Rsf_2=\max_{\Uc_1\subseteq [\Ksf]:|\Uc_1|\geq \Usf} \max_{i\in \Uc_1}  \frac{|Y_i^{\mathcal{U}_1}|}{\Lsf},
\end{align}
where $|\cdot|$ represents  the number of symbols inside.
 The objective of the considered problem is to characterize the capacity region of the rate tuples $(\Rsf_1,\Rsf_2)$, by designing the keys $(P_i)_{i\in [\Ksf]}$, the queries $(Q_{1,i})_{i\in [\Ksf]}$,  $(Q^{\Uc_1}_{2,i})_{i\in\Uc_1}$, the two-round transmissions $(X_i)_{i\in[\Ksf]}$ and $(Y_{i}^{\mathcal{U}_1})_{i\in\mathcal{U}_1}$ for any $\Uc_1 \subseteq [\Ksf]$ with $|\Uc_1|\geq \Usf$, satisfying the constraints in~\eqref{eq:decodability}-\eqref{eq:privacy}. 

\subsection{Related Results}
The optimal rate region of the information theoretic secure aggregation problem against user dropouts and collusion was proposed in \cite{so2022lightsecagg}, which focuses on $\sf K_c=1$ linear combination of users' inputs.
\begin{thm}[\cite{so2022lightsecagg}]
\label{lightsecagg:converse}
    For the information theoretic secure aggregation problem with $\Ksf$ users and  at least $\Usf$ non-dropped users, where $\sf 1\leq U\leq {K-1}$, the capacity region is
   \begin{equation}
      \mathcal{\Rsf}^{\star}=\left\{\left(\Rsf_1,\Rsf_2\right):\Rsf_1\geq 1, \Rsf_2\geq \frac{1}{\Usf}\right\}. 
   \end{equation}
\end{thm}

We then consider on the symmetric PC problem and review the following scheme which will be used later.
\begin{thm}[\cite{zhu2022symmetric}]
\label{symmetricpc}
    For the symmetric PC problem with $\Nsf$ databases and $\Ksf$ files, where there are at least $\Usf>1$ non-dropped databases and the computation task is one linear combination of the $\Ksf$ files,  the following rate   is achievable, 
    \begin{align}
        \sf R_{PC}=1-\frac{1}{U}.
        \label{eq:sPC scheme}
    \end{align}
\end{thm}

\section{Main Results}
For the $(\Ksf,\Usf,{\sf K_c})$ multi-message secure aggregation with demand privacy problem, this section introduces the main results of this paper, including one scheme for the case ${\sf K_c}=1$ proven to be exactly optimal, and another scheme for the case $2\leq {\sf K_c}< \Usf$ proven to be order optimal within $2$. 

\begin{thm}
\label{thm: first result}
   For the $(\Ksf,\Usf,{\sf K_c})$ multi-message secure aggregation with demand privacy problem  where $ \Usf\leq \Ksf-1$ and  ${\sf K_c}=1$, the capacity region  is
   \begin{equation}\label{eq: first result}
      \mathcal{R}^{\star}=\left\{\sf \left(R_1,R_2\right):R_1\geq 1, R_2\geq \frac{1}{U}\right\}. 
   \end{equation}
\end{thm}

The converse bound for Theorem~\ref{thm: first result} can be directly derived from Theorem~\ref{lightsecagg:converse}. For the achievability, we propose a secure aggregation scheme in Section~\ref{sec: proof of first result}, by combining the multiplicative encryption on demand and the additive encryption on models. 

For the case of ${\sf K_c} \geq 2$, a baseline approach is to simply repeat the scheme in Theorem~\ref{thm: first result} for $\sf K_c$ times, achieving the rate tuple in the following theorem. 
\begin{thm}
\label{baseline}
    For the $(\Ksf,\Usf,{\sf K_c})$ multi-message secure aggregation with demand privacy problem where $ \Usf\leq \Ksf-1$, the following rate tuple $\left(\Rsf_1 ={\sf K_c},\Rsf_2= \frac{{\sf K_c}}{\Usf} \right)$ is achievable. 
\end{thm}

 To improve the baseline approach, we introduce a new scheme by leveraging the symmetric PC scheme in \cite{zhu2022symmetric}. The achieved rate tuple is described in the following theorem, whose achievability could be found in Section~\ref{sec: proof of second result}.
\begin{thm}
\label{thm: second result}
    For the $(\Ksf,\Usf,{\sf K_c})$ multi-message secure aggregation with demand privacy problem where  $2\leq {\sf K_c}< \Usf\leq \Ksf-1$, the
      rate tuple $\left( \Rsf_1= 1,\Rsf_2= \frac{{\sf K_c}}{\Usf-1} \right)$ is achievable, 
\end{thm}

Compared to the baseline scheme, the proposed scheme in Theorem~\ref{thm: second result} reduces the first-round rate from     $\sf K_c$ to $1$, while increasing the second-round rate from $\Ksf_{ \csf}/\Usf$ to $\Ksf_{\csf}/(\Usf-1)$. 

The following theorem provides a converse  bound for the considered problem, whose proof could be found in the extended version of this paper~\cite[Appendix A]{privatesecure}.
\begin{thm}
\label{converse of second result}
  For the $(\Ksf,\Usf,{\sf K_c})$ multi-message secure aggregation with demand privacy problem  where $ \Usf\leq \Ksf-1$, any achievable rates should satisfy
   \begin{equation}
      \Rsf_1\geq 1, \ \ \Rsf_2\geq \frac{\Ksf_\csf}{\Usf}. \label{eq:converse for Kc>1}
   \end{equation}
\end{thm}

Comparing  the achievable scheme in Theorem~\ref{thm: second result} and the converse bound in  Theorem~\ref{converse of second result}, 
we can see that the first-round rate of the proposed scheme is exactly optimal and the second-round rate is order optimal within $2$, since 
\begin{align*}
    {\sf \frac{K_c/(U-1)}{K_c/U}=\frac{U}{U-1}} \leq2,
\end{align*}
for any $ \Usf\geq2$. 

\begin{figure}
  \centering
  \includegraphics[width=0.5\textwidth]{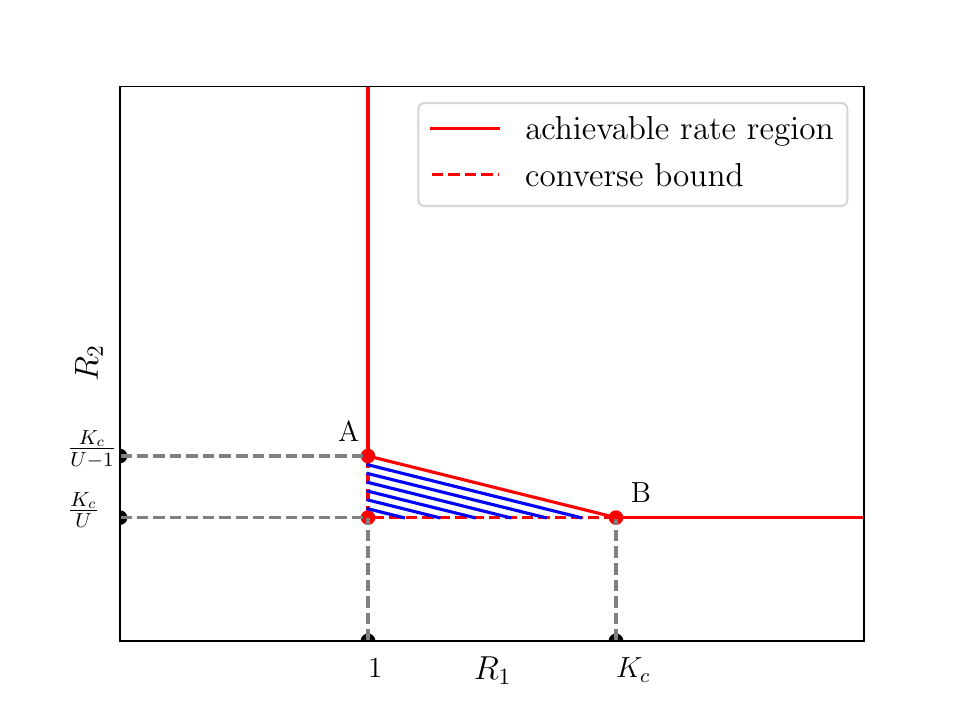}
  \caption{Rate comparison between Theorem~\ref{baseline} and Theorem~\ref{thm: second result}}
  \label{fig: rate comparison}
\end{figure}

In Fig.~\ref{fig: rate comparison} we compare the convex envelope of the achievable bounds in Theorems~\ref{baseline} (Point B) and~\ref{thm: second result} (Point A), with the converse bound in Theorem~\ref{converse of second result}. 
The gap between the achievable and converse bounds is illustrated by the shaded region in Fig.~\ref{fig: rate comparison}.

\section{Achievable Scheme for Theorem~\ref{thm: first result}}
\label{sec: proof of first result}
In this section, we propose a secure aggregation scheme with demand privacy for the case ${\sf K_c}=1$. We first present an example to illustrate the main idea.

\subsection{Example: $(\Ksf,\Usf,\Ksf_\csf)=(3,2,1)$}
We consider the case with $\Ksf=3$ users and at least $\Usf=2$ non-dropped users. The desired linear combination of local models is $a_{1,1}W_1+a_{1,2}W_2+a_{1,3}W_3$, for some coefficients $a_{1,1}, a_{1,2}, a_{1,3}$.  The proposed scheme has the following four stages:

{\bf Stage 1 (Query generation).} 
The server randomly and uniformly picks a random variable $t \in \mathbb{F}_{\qsf} \backslash\{0\}$. Define 
\begin{align}
          Q_{1,1}=\frac{1}{ta_{1,1}},  Q_{1,2}=\frac{1}{ta_{1,2}}, Q_{1,3}=\frac{1}{ta_{1,3}}, 
     \label{eq:a b c}
\end{align}
and then the server sends queries 
$Q_{1,1}=\alpha , Q_{1,2}=\beta ,Q_{1,3}=\gamma$
to users $1, 2, 3$, respectively. Note that~\eqref{eq:a b c} holds since none of  $a_{1,1}, a_{1,2}, a_{1,3}$ is $0$, by the problem formulation. 

{\bf Stage 2 (Key generation).} The key generation is the same as the secure aggregation scheme  in~\cite{so2022lightsecagg}. 
For each  $i\in\{1,2,3\}$, we generate $Z_i$ uniformly i.i.d. over $\mathbb{F}_{\qsf}^{\Lsf}$ and divide it into $\Usf=2$ sub-keys.
Then the sub-keys are encoded as
\begin{equation}
   \begin{bmatrix}
      [\tilde{Z}_1]_1 & [\tilde{Z}_1]_2 & [\tilde{Z}_1]_3\\
      [\tilde{Z}_2]_1 & [\tilde{Z}_2]_2 & [\tilde{Z}_2]_3\\
      [\tilde{Z}_3]_1 & [\tilde{Z}_3]_2 & [\tilde{Z}_3]_3
   \end{bmatrix}
   =
   \begin{bmatrix}
      [Z_1]_1 & [Z_1]_2\\
      [Z_2]_1 & [Z_2]_2\\
      [Z_3]_1 & [Z_3]_2
   \end{bmatrix}
   \begin{bmatrix}
      1 & 1 & 1\\
      1 & 2 & 4
   \end{bmatrix}.
\end{equation}
Each user $i$ stores the keys $Z_i$ and $[\tilde{Z}_j]_i$ for $j \in [3] \setminus \{i\}$, i.e., we have 
$P_i=(Z_i, ([\tilde{Z}_j]_i :j \in [3] \setminus \{i\} ) )$ for each $i\in[3]$.


{\bf Stage 3 (First-round transmission).} Users mask  their local models  as follows:
\begin{equation}
\label{example:first round message}
      X_1=W_1+\alpha Z_1,\quad X_2=W_2+\beta Z_2,\quad X_3=W_3+\gamma Z_3,
\end{equation}
and send its encrypted model to the server. 

Suppose that user $3$ drops in the previous stage, $\mathcal{U}_1=\{1,2\}$ and the server wants to compute
$a_{1,1}W_1+a_{1,2}W_2$.

{\bf Stage 4 (Second-round transmission).} The objective of the second-round is to let the server recover  $Z_1+Z_2$. The server first informs the surviving users $1$ and $2$ of $\mathcal{U}_1=\{1,2\}$.  Users $1$ and $2$ then send $Y_{1}^{\{1,2\}}$ and $Y_{2}^{\{1,2\}}$:
\begin{align}
\label{example:second round message1}
   Y_{1}^{\{1,2\}}&=[\tilde{Z}_1]_1+[\tilde{Z}_2]_1\notag \\
   &=([Z_1]_1+[Z_2]_1)+([Z_1]_2+[Z_2]_2),\\
   \label{example:second round message2}
   Y_{2}^{\{1,2\}}&=[\tilde{Z}_1]_2+[\tilde{Z}_2]_2\notag \\
   &=([Z_1]_1+[Z_2]_1)+2([Z_1]_2+[Z_2]_2),
\end{align}
to the server respectively. Note that there should be at least $2$ users surviving, thus no user drops in the second round. After receiving $ Y_{1}^{\{1,2\}}$ and $Y_{2}^{\{1,2\}}$, the server can decode $Z_1+Z_2$. Note that in this scheme, the server does not need to send queries in the second round.

From the messages received over the first round, the server can obtain
\begin{equation}
   \frac{1}{\alpha}X_1+\frac{1}{\beta}X_2 = \frac{1}{\alpha}W_1+\frac{1}{\beta}W_2+(Z_1+Z_2).
\end{equation}
$Z_1+Z_2$ is decoded in the procedure above. 
By the above selection of $t, \alpha, \beta$, we have 
\begin{equation}
   t(a_{1,1}W_1+a_{1,2}W_2)=\frac{1}{\alpha}W_1+\frac{1}{\beta}W_2.
   \label{eq:solving alpha}
\end{equation}
Hence,  the server can decode $a_{1,1}W_1+a_{1,2}W_2$ and satisfy the decodability constraint.

By the above method, the achieved rates are $\Rsf_1=1$ and $\Rsf_2=1/2$, coinciding with Theorem~\ref{thm: first result}.


{\bf Proof of Security.} We provide an intuitive proof on the security constraint of the proposed scheme.

In the first round of transmission, the messages are
\begin{equation}
   \begin{bmatrix}
      \frac{1}{\alpha} X_1\\
      \frac{1}{\beta} X_2\\
      \frac{1}{\gamma} X_3
   \end{bmatrix}
   =
   \begin{bmatrix}
      \frac{1}{\alpha} & 0 &  0 & 1 & 0 & 0\\
      0 & \frac{1}{\beta} & 0 & 0 & 1 & 0\\
      0 & 0 & \frac{1}{\gamma} & 0 & 0 &  1 
   \end{bmatrix}
   \begin{bmatrix}
      W_1\\
      W_2\\
      W_3\\
      Z_{1}\\
      Z_{2}\\
      Z_{3}
   \end{bmatrix}.
\end{equation}

Consider the coefficient matrix of the keys $Z_{1},Z_{2},Z_{3}$, any submatrix with dimension $|\mathcal{U}_1|\times|\mathcal{U}_1|$, where $|\mathcal{U}_1|\geq2$, is full rank.
As a result, the keys can fully mask the local models in the first round.
In the second round, the server can recover the sum of the keys $Z_1+Z_2$ containing $\Lsf$ symbols, by the seminal information theoretic security results by Shannon~\cite{shannonsecurity}, the server can at most recover $\Lsf$ symbols about the inputs, which is $a_{1,1}W_1+a_{1,2}W_2$.
Thus the security constraint is satisfied. 

{\bf Proof of Privacy.}
Recall that $\alpha=\frac{1}{ta_{1,1}}$ and $ \beta=\frac{1}{ta_{1,2}}$.
Since $a_{1,i}\neq 0$ for each $i\in[3]$ and $t$ is a random variable from $\mathbb{F}_\qsf \backslash\{0\}$, $\alpha$ is independent of $\frac{1}{a_{1,1}}$ due to the multiplicative cipher. Consequently, user $1$ cannot infer anything about the desired linear combination coefficient $a_{1,1}$ from $\alpha$. The same holds for user $2$, ensuring that the privacy constraint is satisfied.

Similarly, the proposed scheme works for any $\Uc_1$ and $\Uc_2$, where $\Uc_2\subseteq  \Uc_1 \subseteq [\Ksf]$ and  $|\Uc_2|\geq \Usf$. 

\subsection{General Scheme}
We now present the general secure aggregation scheme for Theorem~\ref{thm: first result}.

{\bf Stage 1 (Query generation).} 
The server picks a random variable $t \in \mathbb{F}_{\qsf} \backslash\{0\}$ and we define
\begin{equation}
\label{eq:q,1,i}
   Q_{1,i}=\frac{1}{ta_{1,i}}, i\in[\Ksf].
\end{equation}
The query $Q_{1,i}$ is then sent to user $i$. Note that~\eqref{eq:q,1,i} holds because $(a_{1,i}:i\in[\Ksf])$ are non-zero elements, as specified in the problem formulation.

{\bf Stage 2 (Key generation).} According to \cite{so2022lightsecagg},we generate $Z_i$ uniformly i.i.d. over $\mathbb{F}_{\qsf}^{\Lsf}$ and divide each $Z_i$ into $\Usf$ sub-keys, i.e. $[Z_i]_m\in \mathbb{F}_{\qsf}^{\frac{\Lsf}{\Usf}}, m\in [\Usf]$. Then user $i\in [\Ksf]$ encodes sub-keys as 
\begin{equation}
   [\tilde{Z}_{i}]_j=\left([Z_i]_1,\ldots,[Z_i]_{\Usf}\right)\cdot \mathbf{M}_{:,j},
\end{equation}
where $\mathbf{M}_{:,j}$ is $j$-th column of a MDS matrix $\mathbf{M}\in \mathbb{F}_{\qsf}^{\sf U\times K}$. Each user $i$ eventually stores $P_i=(Z_i, ([\tilde{Z}_j]_i :j \in [\Ksf] \setminus \{i\} ) )$.

{\bf Stage 3 (First-round transmission).} User $i$ masks its local models as 
\begin{equation}
\label{first round message}
   X_i=W_i+Q_{1,i}Z_{i},
\end{equation}
and sends it to the server. The server receives the messages from users in $\mathcal{U}_1$, where $|\mathcal{U}_1|\geq\Usf$.
After receiving the messages from two rounds, the server wants to recover $\sum_{i\in{\mathcal{U}_1}}a_{1,i}W_i$.

{\bf Stage 4 (Second-round transmission).} To recover $\sum_{i\in{\mathcal{U}_1}}{Z_{i}}$ for decryption, the server informs the surviving users of $\mathcal{U}_1$ and user $j\in \mathcal{U}_1$ sends its aggregated encoded sub-keys $\sum_{i\in{\mathcal{U}_1}}[\tilde{Z}_{i}]_j$ to the server. Note that $\sum_{i\in{\mathcal{U}_1}}[\tilde{Z}_{i}]_j$ is an encoded version of $\sum_{i\in{\mathcal{U}_1}}[Z_{i}]_m$ for $m\in [\Usf]$ using the MDS matrix $\mathbf{M}$. 
Thus, after receiving a set of any $\Usf$ messages from the surviving users, the server is able to decode 
$\sum_{i\in{\mathcal{U}_1}}{Z_{i}}$ by concatenating $\sum_{i\in{\mathcal{U}_1}}[Z_{i}]_m$'s. Note that in this scheme, the server does not need to send queries in the second round.

During the first-round transmission, the server can obtain
\begin{equation}
   \sum_{i\in{\mathcal{U}_1}}\frac{1}{Q_{1,i}}X_i=\sum_{i\in{\mathcal{U}_1}}\frac{1}{Q_{1,i}}W_i+\sum_{i\in{\mathcal{U}_1}}Z_{i},
\end{equation}
where $\sum_{i\in{\mathcal{U}_1}}Z_{i}$ is recovered in the second-round transmission. According to the design of $t$ and $(Q_{1,i})_{i\in{\mathcal{U}_1}}$, we have
\begin{equation}
   t\sum_{i\in\mathcal{U}_1}a_{1,i}W_i=\sum_{i\in\mathcal{U}_1}\frac{1}{Q_{1,i}}W_i.
\end{equation}
In this way, the server is able to decode $\sum_{i\in{\mathcal{U}_1}}a_{1,i}W_i$ and satisfy the decodability constraint.

The above scheme has achieved $\sf R_1=1$ and $\sf R_2=\frac{L/U}{L}=\frac{1}{U}$, for the aggregated sub-keys $\sum_{i\in{\mathcal{U}_1}}[\tilde{Z}_{i}]_j$ sent in second-round transmission has the length $\sf L/U$. The result coincides with Theorem~\ref{thm: first result}.

The proof of security and privacy of the above scheme could be found in~\cite[Appendix B]{privatesecure}.

\section{Achievable Scheme for Theorem~\ref{thm: second result}}
\label{sec: proof of second result}
When the server requests $\sf K_c\geq 2$ linear combinations of users' inputs, we introduce robust symmetric private computation in the second-round transmission to guarantee the constraints of both security and privacy. The following three stages are included in our proposed scheme. Due to the limitation of pages, the following content illustrates the main idea, while the detailed description could be found in ~\cite[Appendix C]{privatesecure}.




{\bf Stage 1 (Key generation).} For each $i\in [\Ksf]$, we pick $Z_i$ from $\mathbb{F}_{\qsf}^{\Lsf}$ uniformly and randomly, and share  
$Z_i$ to all users. Thus    $P_i=(Z_{i})_{i\in[\Ksf]}$. Users also share $\Lsf/(\Usf-1)$ common random variables distributed uniformly over $\mathbb{F}_\qsf$ as the mask of keys.

{\bf Stage 2 (First-round transmission).} Users  mask their local models as 
\begin{equation}
    X_i=W_i+Z_i,
\end{equation}
and send its encrypted model to the server. Note that the server does not need to send queries in the first round. After this round, the server wishes to compute $\left(\sum_{i\in\mathcal{U}_1}a_{n,i}W_i\right)_{n\in[\sf K_c]}$ securely and privately.


{\bf Stage 3 (Second-round transmission).}  
From the messages received in the first round, the server obtains
\begin{equation}
    \sum_{i\in \mathcal{U}_1} a_{n,i}X_i=(\sum_{i\in \mathcal{U}_1} a_{n,i}W_i)+(\sum_{i\in \mathcal{U}_1} a_{n,i}Z_{i}).
\end{equation}

Hence, $\sf K_c$ combinations of keys $(Z_{i})_{i\in\mathcal{U}_1}$ must be recovered from the second-round transmission for decryption. We have found that the coefficients of the keys' combinations can leak information about the server's computation task. Thus, to ensure privacy and security of the scheme, we apply the symmetric PC scheme in~\cite{zhu2022symmetric} $\sf K_c$ times to obtain $\sf K_c$ combinations of keys separately.

We consider keys $(Z_{i})_{i\in[\Ksf]}$ as the messages in symmetric PC and divide each key $Z_i$ into multiple non-overlapping and equal-length sub-keys. One sub-key $\tilde{Z}_i$ contains $\Lsf^{'}=\Usf-1$ symbols.
A linear combination of sub-keys can be represented as a dependent message 
\begin{align}
  V^{\theta_n}&={\phi}^{\theta_n}(\tilde{Z}_1,\cdots,\tilde{Z}_{\Ksf})\notag \\
  &=\sum_{i\in \mathcal{U}_1} a_{n,i}\tilde{Z}_{i}.
\end{align}
${\theta}_n$ is the index of the $n$-th  desired combination among all the linear combinations of $\tilde{Z}_{1},\cdots,\tilde{Z}_{\Ksf}$, where $ n\in [\Ksf_\csf]$.

During each retrieval, the server sends a query $Q^{\Uc_1}_{2,i}$ to each user in $\Uc_1$ and receives an one-symbol answer from each non-dropped user, in order to recover a linear combination of sub-keys. By concatenating the results $\sum_{i\in \mathcal{U}_1} a_{n,i}\tilde{Z}_{i}$ for all sub-keys, the server can recover $\sum_{i\in \mathcal{U}_1} a_{n,i}Z_{i}$. 

After the second-round transmission, the server (i) decodes $\left(\sum_{i\in\mathcal{U}_1}a_{n,i}W_i\right)_{n\in[\sf K_c]}$ correctly; (ii) gains no more information about users' keys except for the computation task and protects the identity of its desired combinations from users; (iii) achieves the rate tuple $\left( \Rsf_1= 1,\Rsf_2= \frac{{\Ksf_\csf}}{\Usf-1} \right)$. Each key $Z_i$ contains $\sf L/L^{'}$ sub-keys and thus $\sf R_2=K_c\times \frac{L/L^{'}}{L}=\frac{K_c}{U-1}$, coinciding with Theorem~\ref{thm: second result}.

{\bf Proof of Security.} The intuitive proof of security in the first round followed by the proof in Section~\ref{sec: proof of first result}.
The submatrix of keys is $\sf K\times K$ (resp. $|\mathcal{U}_1|\times|\mathcal{U}_1|$ when dropped users exist) full rank, fully masking the $\Ksf$-dimensional (resp. $|\mathcal{U}_1|$-dimensional) received messages.
In the second round, we apply the symmetric PC 
and only decode $\sf K_c$ dimension of information about $Z_i$'s, subsequently recovering only $\sf K_c$-dimensional information about $W_i$'s, which is exactly the computation task of the server. No more information about users' local models is leaked.

{\bf Proof of Privacy.} Possible leakage of the server's task will only happen in the second round, where we introduce symmetric PC to address this problem. 
By designing the second-round queries $(Q^{\Uc_1}_{2,i})_{i\in\Uc_1}$, the query polynomials are indistinguishable from the perspective of each user and the privacy constraint is satisfied in this way.

\section{Conclusion}
This paper introduces a new problem on multi-message secure aggregation with demand privacy, ensuring the security of users' local inputs and the privacy of the server's computation task. Our proposed schemes allow the server to decode $\Ksf_\csf\geq 1$ linear combinations of users' local inputs, with tolerance against $\sf K-U$ dropped users. For
${\sf K_c}=1$, a secure aggregation with demand privacy scheme was proposed to achieve the optimal rate region; for $2\leq {\sf K_c}< \Usf$, another proposed scheme achieves the optimal first-round rate and order optimal second-round rate (within $2$).
On-going works include the characterization of the capacity region in the open regime, and the minimization of the required key size.


\bibliographystyle{IEEEtran}
\bibliography{ref}


\appendices

\section{Proof of Theorem~\ref{converse of second result}}
\label{appendix: converse of second result}

\subsection{Proof of $\Rsf_1\geq 1$}
We prove the converse bound for the first-round transmission rate. Consider any $u\in [\Ksf]$, and set $\Uc_1=[\Ksf], \Uc_2=[\Ksf]\backslash\{u\}$. From the decodability constraint \eqref{eq:decodability}, we have
\begin{align}
    0=H\Biggl(&\left.\left(\sum_{i\in[\Ksf]}a_{n,i}W_i\right)_{n\in[\sf K_c]}\right|(X_i)_{i\in[\Ksf]}, (Q_{1,i})_{i\in[\Ksf]}, \cdots \notag\\
    &\cdots(Q^{[\Ksf]}_{2,i})_{i\in[\Ksf]}, (Y_{i}^{[\Ksf]})_{i\in[\Ksf]\backslash\{u\}}\Biggr)\\
    \geq H\Biggl(&\left.\left(\sum_{i\in[\Ksf]}a_{n,i}W_i\right)_{n\in[\sf K_c]}\right|(X_i)_{i\in[\Ksf]}, (Q_{1,i})_{i\in[\Ksf]}, \cdots\notag \\
    &\cdots(Q^{[\Ksf]}_{2,i})_{i\in[\Ksf]}, (Y_{i}^{[\Ksf]})_{i\in[\Ksf]\backslash\{u\}}, (W_i,P_i)_{i\in[\Ksf]\backslash\{u\}}\Biggr)\\
    =H\Biggl(&\left.W_u\right|X_u, (Q_{1,i})_{i\in[\Ksf]}, (Q^{[\Ksf]}_{2,i})_{i\in[\Ksf]}, (W_i,P_i)_{i\in[\Ksf]\backslash\{u\}}\Biggr),\label{eq: function}
\end{align}
where \eqref{eq: function} follows from the fact that 
$(X_{i}^{[\Ksf]})_{i\in[\Ksf]\backslash\{u\}}$ and $(Y_{i}^{[\Ksf]})_{i\in[\Ksf]\backslash\{u\}}$ are functions of $(W_i,P_i)_{i\in[\Ksf]\backslash\{u\}}$, $(Q_{1,i})_{i\in[\Ksf]\backslash\{u\}}$ and $(Q^{[\Ksf]}_{2,i})_{i\in[\Ksf]\backslash\{u\}}$. Thus, we have
\begin{align}
    \Lsf&=H\left(W_u\right)\notag \\
    &=H\left(\left.W_u\right|(Q_{1,i})_{i\in[\Ksf]}, (Q^{[\Ksf]}_{2,i})_{i\in[\Ksf]}, (W_i,P_i)_{i\in[\Ksf]\backslash\{u\}}\right)\\
    &=I\left(\left.W_u;X_u\right|(Q_{1,i})_{i\in[\Ksf]}, (Q^{[\Ksf]}_{2,i})_{i\in[\Ksf]}, (W_i,P_i)_{i\in[\Ksf]\backslash\{u\}}\right)\\
    &\leq H\left(\left.X_u\right|(Q_{1,i})_{i\in[\Ksf]}, (Q^{[\Ksf]}_{2,i})_{i\in[\Ksf]}, (W_i,P_i)_{i\in[\Ksf]\backslash\{u\}}\right)\\
    &\leq H(X_u) \Rightarrow \Rsf_1\geq 1.
\end{align}

\subsection{Proof of $\Rsf_2\geq \frac{\Ksf_\csf}{\Usf}$}
We prove the converse bound for the second-round transmission rate. Consider $\Uc_1=[\Usf+1], \Uc_2=[\Usf]$ and from the security constraint \eqref{eq:security}, we have
\begin{align}
    0=I\Biggl(&\left.(W_i)_{i\in[\Ksf]};(X_i)_{i\in[\Ksf]},(Y_{i}^{[\Usf+1]})_{i\in[\Usf+1]}\right|\cdots\notag \\
    &\cdots (\sum_{i\in[\Usf+1]}a_{n,i}W_i)_{n\in[\sf K_c]}\Biggr)\\
    \geq I\Biggl(&\left.(\sum_{i\in[\Usf]}a_{n,i}W_i)_{n\in[\sf K_c]};(X_i)_{i\in[\Usf]}\right|(\sum_{i\in[\Usf+1]}a_{n,i}W_i)_{n\in[\sf K_c]}\Biggr)\\
    =I\Biggl(&(\sum_{i\in[\Usf]}a_{n,i}W_i)_{n\in[\sf K_c]};(X_i)_{i\in[\Usf]}, (\sum_{i\in[\Usf+1]}a_{n,i}W_i)_{n\in[\sf K_c]}\Biggr)\notag \\
    -I&\left((\sum_{i\in[\Usf]}a_{n,i}W_i)_{n\in[\sf K_c]};(\sum_{i\in[\Usf+1]}a_{n,i}W_i)_{n\in[\sf K_c]}\right)\\
    \geq I\Biggl(&(\sum_{i\in[\Usf]}a_{n,i}W_i)_{n\in[\sf K_c]};(X_i)_{i\in[\Usf]}\Biggr).\label{eq:information1}
\end{align}

Now we consider $\Uc_1=\Uc_2=[\Usf]$, From \eqref{eq:decodability}, we have
\begin{align}
    0=H\Biggl(&\left.(\sum_{i\in[\Usf]}a_{n,i}W_i)_{n\in[\sf K_c]}\right|(X_i)_{i\in[\Usf]}, (Q_{1,i})_{i\in[\Usf]}, \cdots \notag\\
    &\cdots(Q^{[\Usf]}_{2,i})_{i\in[\Usf]}, (Y_{i}^{[\Usf]})_{i\in[\Usf]}\Biggr)\\
    \Rightarrow \Ksf_\csf \Lsf= I\Biggl(&(\sum_{i\in[\Usf]}a_{n,i}W_i)_{n\in[\sf K_c]};(X_i)_{i\in[\Usf]}, (Q_{1,i})_{i\in[\Usf]},\cdots\notag \\
    &\cdots (Q^{[\Usf]}_{2,i})_{i\in[\Usf]}, (Y_{i}^{[\Usf]})_{i\in[\Usf]}\Biggr)\\
    =I\Biggl(&(\sum_{i\in[\Usf]}a_{n,i}W_i)_{n\in[\sf K_c]};(Q_{1,i})_{i\in[\Usf]},\cdots \notag\\
    &\cdots \left.(Q^{[\Usf]}_{2,i})_{i\in[\Usf]}, (Y_{i}^{[\Usf]})_{i\in[\Usf]}\right|(X_i)_{i\in[\Usf]}\Biggr) \label{eq:information2}\\ 
    \leq H\Biggl(&(Y_{i}^{[\Usf]})_{i\in[\Usf]}\Biggr) \leq \sum_{i\in[\Usf]}H\left(Y_{i}^{[\Usf]}\right)\leq\Usf\Lsf_\Ysf\\
    \Rightarrow \Rsf_2\geq \frac{\Ksf_\csf}{\Usf},
\end{align}
where \eqref{eq:information2} follows from \eqref{eq:information1}.

\section{Proof of Security and Privacy in Section~\ref{sec: proof of first result}}
\label{appendix: first proof}

\subsection{Proof of Security}
We provide an intuitive proof on the security constraint of the general scheme in Section~\ref{sec: proof of first result}. Consider 

Over the first-round transmission, the messages can be denoted as a matrix as follows:
\begin{small}
\begin{equation}
   \begin{bmatrix}
      \frac{1}{Q_1} X_1\\
      \vdots\\
      \frac{1}{Q_K} X_\Ksf
   \end{bmatrix}
   =
   \begin{bmatrix}
      \frac{1}{Q_1} & 0 & \cdots & 0 & 1 & 0 & \cdots & 0\\
      \vdots & \vdots & \ddots & \vdots & \vdots & \vdots & \ddots & \vdots\\
      0 & 0 & \cdots & \frac{1}{Q_\Ksf} & 0 & 0 & \cdots & 1 
   \end{bmatrix}
   \begin{bmatrix}
      W_1\\
      \vdots\\
      W_\Ksf\\
      Z_{1}\\
      \vdots\\
      Z_{\Ksf}
   \end{bmatrix}
\end{equation}
\end{small}

The coefficient matrix of the keys $Z_{1},\ldots,Z_{\Ksf}$, any submatrix with dimension $|\mathcal{U}_1|\times|\mathcal{U}_1|$, where $|\mathcal{U}_1|\geq \Usf$, is full rank. Thus, the keys can fully mask the $|\mathcal{U}_1|$-dimensional local models in the first round.

Over the second-round transmission, the server computes the sum of the keys $\sum_{i\in{\mathcal{U}_1}}{Z_{i}}$, containing $\Lsf$ symbols. 
Combined with the previous received messages, the server can at most recover $\Lsf$ symbols about users' local inputs according to seminal information theoretic security results in \cite{shannonsecurity}s, 
which is $\sum_{i\in{\mathcal{U}_1}}\frac{1}{Q_i}{W_i}$.
No more information can be obtained by the server and consequently, the second round satisfies the security constraint.

\subsection{Proof of Privacy}
In order to meet the correctness constraint, we introduce a random variable $t$ so that the definition of $Q_i, i\in\mathcal{U}_1$ satisfies \eqref{eq:q,1,i}.  
The values of $Q_i$ is decoded as a function of $a_{1,i}$ and $t$, 
and since $a_{1,i}\neq 0$ and $t$ is a random variable from $\mathbb{F}_\qsf \backslash\{0\}$ with equal probability distribution, $Q_i$ is independent of $a_{1,i}, i\in{\mathcal{U}_1}$ according to the multiplicative cipher. Privacy constraint is satisfied.

\section{Detailed Achievable Scheme For Theorem~\ref{thm: second result}}
\label{appendix: detailed second scheme}

\subsection{Example: $(\Ksf,\Usf,\Ksf_\csf)=(4,3,2)$}

{\bf Stage 1 (Key generation).} For each $i\in [4]$, we pick $Z_i$ from $\mathbb{F}_{\qsf}^{\Lsf}$, uniformly at random, and share  
$Z_i$ to all users. Thus $P_1=P_2=P_3=P_4=(Z_{i})_{i\in[4]}$. Additional $\sf L/(U-1)=L/2$ random variables distributed uniformly over $\mathbb{F}_\qsf$ are shared among users as the mask of keys.

{\bf Stage 2 (First-round transmission).} Users $i\in [4]$ mask their local models as 
\begin{equation}
    \begin{split}
        X_1=W_1+Z_{1}, \quad &X_2=W_2+Z_{2},\\
        X_3=W_3+Z_{3}, \quad &X_4=W_4+Z_{4},
    \end{split}
\end{equation}
and send its encrypted model to the server. 

{\bf Stage 3 (Second-round transmission).} Assume now $\Uc_1=[4]$, i.e., no user drops in the first round. 
From the messages received over the first round, the server obtains
\begin{align}
    \sum_{i\in[4]}X_i=&\sum_{i\in[4]}W_i+\sum_{i\in[4]}Z_i,\\
    \sum_{i\in[4]}iX_i=&\sum_{i\in[4]}iW_i+\sum_{i\in[4]}iZ_i.
\end{align}

Thus, $K_c=2$ combinations of keys $(Z_{i})_{i\in[4]}$ must be recovered from the second-round transmission for decryption. 
To ensure privacy and security of the scheme, we apply the symmetric PC scheme in~\cite{zhu2022symmetric} to retrieve $K_c=2$ combinations of keys separately.

We consider keys $(Z_{i})_{i\in[4]}$ as the messages in symmetric PC and divide each key $Z_i$ into multiple non-overlapping and equal-length sub-keys. One sub-key $\tilde{Z}_i$ contains $\Lsf^{'}=\Usf-1=2$ symbols.
A linear combination of sub-keys can be represented as a dependent message, \begin{align}
    &V^{\theta_1}={\phi}^{\theta_1}(\tilde{Z}_1,\cdots,\tilde{Z}_4)=\tilde{Z}_1+\tilde{Z}_2+\tilde{Z}_3+\tilde{Z}_4,\\
    &V^{\theta_2}={\phi}^{\theta_2}(\tilde{Z}_1,\cdots,\tilde{Z}_4)=\tilde{Z}_1+2\tilde{Z}_2+3\tilde{Z}_3+4\tilde{Z}_4.
\end{align}

To retrieve $V^{\theta_1}$ of length $\sf L^{'}=2$, the server independently and uniformly generates 2 random linear functions $\phi_1, \phi_2$ from the vector space $\mathcal{P}$ containing all possible linear combinations of $(\tilde{Z}_{i})_{i\in[4]}$. Also for the construction of queries, distinct elements $(\alpha_i)_{i\in[4]}$ and $\beta_1, \beta_2$ are selected to apply Lagrange encoding. For each $i\in[2]$, generate the query polynomial function $\rho_i(x_1,\cdots,x_\Ksf,\alpha)$ of degree $\sf L^{'}=2$ in variable $\alpha$ such that for $i=1$, 
\begin{align}
    &\rho_1(x_1,\cdots,x_4,\beta_l)=\begin{cases}
        {\phi}^{\theta_1}(x_1,\cdots,x_4) &\text{if}\quad l=1,\\
        0 &\text{if}\quad l=2.
    \end{cases}\\
    &\rho_1(x_1,\cdots,x_4,\alpha_1)=\phi_1(x_1,\cdots,x_4).
\end{align}
So we have
\begin{align}
    \rho_1(x_1,\cdots,x_4,\alpha)&=\phi_1(x_1,\cdots,x_4)\frac{\alpha-\beta_1}{\alpha_1-\beta_1}\frac{\alpha-\beta_2}{\alpha_1-\beta_2}\notag \\
    &+{\phi}^{\theta_1}(x_1,\cdots,x_4)\frac{\alpha-\beta_2}{\beta_1-\beta_2}\frac{\alpha-\alpha_1}{\beta_1-\alpha_1}.
\end{align}
We also construct the query polynomial function for $i=2$ 
\begin{align}
    \rho_2(x_1,\cdots,x_4,\alpha)&=\phi_2(x_1,\cdots,x_4)\frac{\alpha-\beta_1}{\alpha_1-\beta_1}\frac{\alpha-\beta_2}{\alpha_1-\beta_2}\notag \\
    &+{\phi}^{\theta_1}(x_1,\cdots,x_4)\frac{\alpha-\beta_1}{\beta_2-\beta_1}\frac{\alpha-\alpha_1}{\beta_2-\alpha_1}.
\end{align}

The server sends query $Q_{2,i}^{\mathcal{U}_1}$ to user $i$, which is 
\begin{equation}
    Q_{2,i}^{\mathcal{U}_1}=(\rho_1(x_1,\cdots,x_\Ksf,\alpha_i),\rho_2(x_1,\cdots,x_\Ksf,\alpha)), i\in[4].
\end{equation}

After receiving $Q_{2,i}^{\mathcal{U}_1}$, user $i\in[4]$ will send the answer based on data it stored. Let $s$ be a random variable distributed uniformly over $\mathbb{F}_\qsf$ and define an interpolation polynomial $\psi(\alpha)$ of degree $\sf L^{'}=2$, which satisfies
\begin{equation}
    \psi(\beta_l)=0, l=1,2,
\end{equation}
\begin{equation}
    \psi(\alpha_1)=s.
\end{equation}
Then we have
\begin{equation}
    \psi(\alpha)=s\frac{\alpha-\beta_1}{\alpha_1-\beta_1}\frac{\alpha-\beta_2}{\alpha_1-\beta_2}.
\end{equation}

The expression of the answer for all users is $\zeta(\alpha)=\rho_1(\tilde{Z}_{1,1},\cdots,\tilde{Z}_{4,1},\alpha)+\rho_2(\tilde{Z}_{1,2},\cdots,\tilde{Z}_{4,2},\alpha)+\psi(\alpha)$, a 1-symbol message, 
and the user $i\in[4]$ sends its answer by evaluating $\zeta(\alpha)$ at $\alpha={\alpha}_i$, which is $A_i=\zeta(\alpha_i)$. 

$(A_1,\cdots, A_4)=(\zeta(\alpha_1),\cdots,\zeta(\alpha_4))$ forms an $(4,3)$ Reed-Solomon code \cite{raviv2019private}, tolerating $1$ user dropout. Therefore, the server can decode $\zeta(\alpha)$ by at least $\sf U=3$ answers and compute the desired function $V^{\theta_1}$ from
\begin{align}
    \zeta(\beta_1)&=\sum_{\rho=1,2}{\rho_i(\tilde{Z}_{1,i},\cdots,\tilde{Z}_{4,i},\beta_1)}+\psi(\beta_1)\notag \\
    &={\phi}^{\theta_1}(\tilde{Z}_{1,1},\cdots,\tilde{Z}_{4,1}),\\
    \zeta(\beta_2)&=\sum_{\rho=1,2}{\rho_i(\tilde{Z}_{1,i},\cdots,\tilde{Z}_{4,i},\beta_2)}+\psi(\beta_2)\notag \\
    &={\phi}^{\theta_1}(\tilde{Z}_{1,2},\cdots,\tilde{Z}_{4,2}),
\end{align}
where ${\phi}^{\theta_1}(\tilde{Z}_{1,1},\cdots,\tilde{Z}_{4,1})$ and ${\phi}^{\theta_1}(\tilde{Z}_{1,2},\cdots,\tilde{Z}_{4,2})$ are the $\sf L^{'}=2$ elements of $V^{\theta_1}$. By concatenating the results $\sum_{i\in[4]}\tilde{Z}_i$ for all sub-keys, the server can recover $\sum_{i\in[4]}Z_i$.

Using the above scheme again, the server can also decode $V^{\theta_2}$. Hence, the total number of symbols in the second round is $K_c\times \frac{L}{L^{'}}=L$ and $\sf R_2=1$. The first round message $X_i$ contains $\sf L$ symbols and the rate is $\sf R_1=L/L=1$.

\subsection{General Scheme}
Compared to Section~\ref{sec: proof of second result}, this section mainly details the third stage of the general scheme, and thus omits the description of the first and second stages for simplicity. 

{\bf Stage 3 (Second-round transmission).}  
From the messages received in the first round, the server obtains
\begin{equation}
    \sum_{i\in \mathcal{U}_1} a_{n,i}X_i=(\sum_{i\in \mathcal{U}_1} a_{n,i}W_i)+(\sum_{i\in \mathcal{U}_1} a_{n,i}Z_{i}).
\end{equation}
Hence, $\sf K_c$ combinations of keys $(Z_{i})_{i\in\mathcal{U}_1}$ must be recovered from the second-round transmission for decryption. We have found that the coefficients of the keys' combinations can leak information about the server's computation task. Thus, to ensure privacy and security of the scheme, we apply the symmetric PC scheme in~\cite{zhu2022symmetric} $\sf K_c$ times to obtain $\sf K_c$ combinations of keys separately.

We consider keys $(Z_{i})_{i\in[\Ksf]}$ as the messages in symmetric PC and divide each key $Z_i$ into multiple non-overlapping and equal-length sub-keys. One sub-key $\tilde{Z}_i$ contains $\Lsf^{'}=\Usf-1$ symbols.
A linear combination of sub-keys can be represented as a dependent message 
\begin{align}
  V^{\theta_n}&={\phi}^{\theta_n}(\tilde{Z}_1,\cdots,\tilde{Z}_{\Ksf})\notag \\
  &=\sum_{i\in \mathcal{U}_1} a_{n,i}\tilde{Z}_{i}.
\end{align}
${\theta}_n$ is the index of the $n$-th  desired combination among all the linear combinations of $\tilde{Z}_{1},\cdots,\tilde{Z}_{\Ksf}$, where $ n\in [\Ksf_\csf]$.

During each retrieval, the server sends a query $Q^{\Uc_1}_{2,i}$ to each user in $\Uc_1$. For the construction of queries, distinct elements $({\alpha}_n)_{n\in[\Nsf]}$ and $(\beta_l)_{l\in[\Lsf]}$ from $\mathbb{F}_\qsf$ are generated to apply Lagrange encoding. The server also selects $\sf L^{'}$ random linear functions $(\phi_i)_{i\in[\sf L^{'}]}$ independently and uniformly from $\mathcal{P}$, the vector space containing all possible linear combinations of messages, in order to satisfy the privacy constraint. For each $i\in[\sf L^{'}]$, the query polynomial $\rho_i(x_1,\cdots,x_\Ksf,\alpha)$ of degree $\sf L^{'}$ in variable $\alpha$ satisfies
\begin{align}
    &\rho_i(x_1,\cdots,x_\Ksf,\beta_l)=\begin{cases}
        {\phi}^{\theta_n}(x_1,\cdots,x_\Ksf) &\text{if}\quad l=i,\\
        0 &\text{otherwise}
    \end{cases}\\
    &\rho_i(x_1,\cdots,x_\Ksf,\alpha_1)=\phi_i(x_1,\cdots,x_\Ksf)
\end{align}

Thus, $\rho_i(x_1,\cdots,x_\Ksf,\alpha)$ is denoted as
\begin{align}
    &\rho_i(x_1,\cdots,x_\Ksf,\alpha)=\phi_i(x_1,\cdots,x_\Ksf)\times (\prod_{l\in[\sf L^{'}]}\frac{\alpha-\beta_l}{\alpha_1-\beta_l})\notag \\
    &+{\phi}^{\theta_n}(x_1,\cdots,x_\Ksf)\times (\prod_{l\in[\Lsf^{'}],l\neq i}\frac{\alpha-\beta_l}{\beta_i-\beta_l})(\frac{\alpha-\alpha_1}{\beta_i-\alpha_1})
\end{align}
And the query sent to user $i\in\mathcal{U}_1$ is
\begin{equation}
\label{query polynomials}
    Q^{\Uc_1}_{2,i}=(\rho_1(x_1,\cdots,x_\Ksf,\alpha_i),\cdots,\rho_{\Lsf^{'}}(x_1,\cdots,x_\Ksf,\alpha_i))
\end{equation}
which is a linear combination of $\phi^{{\theta}_n}$ and $\phi_i$, and consequently also a function belong to $\mathcal{P}$. Due to the existence of the random function $\phi_i$, the server can protect its demand privacy.

For the generation of users' answers, all users share a random variable $s$ distributed independently and uniformly over $\mathbb{F}_\qsf$ to ensure security. Let an interpolation polynomial $\psi(\alpha)$ of degree $\Lsf^{'}$ satisfy
\begin{equation}
    \psi(\beta_l)=0, \forall l\in[\Lsf^{'}]
\end{equation}
\begin{equation}
    \psi(\alpha_1)=s
\end{equation}
Thus, $\psi(\alpha)$ can be represented as
\begin{equation}
\label{interpolation polynomial}
    \psi(\alpha)=s\prod_{l\in[\Lsf^{'}]}\frac{\alpha-\beta_l}{\alpha_1-\beta_l}
\end{equation}

The expression of the answer for all the surviving users is
\begin{align}
  \zeta(\alpha)&=\sum_{i=1}^{\Lsf^{'}}\rho_i(\tilde{Z}_{1,i},\ldots,\tilde{Z}_{\Ksf,i},\alpha)+\psi(\alpha)\\
  &=\sum_{i=1}^{\Lsf^{'}}\phi_i(\tilde{Z}_{1,i},\ldots,\tilde{Z}_{\Ksf,i})\times (\prod_{l\in[\sf L^{'}]}\frac{\alpha-\beta_l}{\alpha_1-\beta_l})\notag \\
  &+\sum_{i=1}^{\Lsf^{'}}{\phi}^{\theta_n}(\tilde{Z}_{1,i},\ldots,\tilde{Z}_{\Ksf,i})\times (\prod_{l\in[\Lsf^{'}],l\neq i}\frac{\alpha-\beta_l}{\beta_i-\beta_l})(\frac{\alpha-\alpha_1}{\beta_i-\alpha_1})\notag \\
  &+\psi(\alpha)
\end{align}
and the user $i\in \mathcal{U}_1$ sends its answer by evaluating $\zeta(\alpha)$ at $\alpha={\alpha}_i$, which is $A_i=\zeta(\alpha_i)$. 
Possible answers $(\zeta(\alpha_i))_{i\in \mathcal{U}_1}$ form a $(|\mathcal{U}_1|,\Usf)$ RS code \cite{raviv2019private}. The server can recover $\zeta(\alpha)$ by at least $\Usf$ answers from the second round of communication.

For each $l\in[\Lsf^{'}]$, the server calculate $\zeta(\beta_l)$, which is 
\begin{align}
    \zeta(\beta_l)&=\sum_{i=1}^{\Lsf^{'}}\rho_i(Z^i_1,\cdots,Z^i_\Ksf,\beta_l)\\
    &={\phi}^{\theta_n}(Z^l_1,\cdots,Z^l_\Ksf)=V^{\theta_n}_l,
\end{align}
where $V^{\theta_n}_l$ is the $l$-th element of $V^{\theta_n}$. In this way, the server is able to retrieve $\sf L^{'}$ elements of $V^{\theta}$. By concatenating the results $\sum_{i\in \mathcal{U}_1} a_{n,i}\tilde{Z}_{i}$ for all sub-keys, the server can recover $\sum_{i\in \mathcal{U}_1} a_{n,i}Z_{i}$. 

After the second-round transmission, the server (i) decodes $\left(\sum_{i\in\mathcal{U}_1}a_{n,i}W_i\right)_{n\in[\sf K_c]}$ correctly; (ii) gains no more information about users' keys except for the computation task and protects the identity of its desired combinations from users; (iii) achieves the rate tuple $\left( \Rsf_1= 1,\Rsf_2= \frac{{\Ksf_\csf}}{\Usf-1} \right)$. Each key $Z_i$ contains $\sf L/L^{'}$ sub-keys and thus $\sf R_2=K_c\times \frac{L/L^{'}}{L}=\frac{K_c}{U-1}$, coinciding with Theorem~\ref{thm: second result}.







\end{document}